\documentclass[twocolumn,aps,preprint,letter,10pt]{revtex4}
\usepackage{graphicx,graphics,float}
\usepackage{dcolumn}
\usepackage{bm}
\usepackage{amsmath,amssymb}
\usepackage{mathrsfs}
\usepackage{xcolor}
\begin{document}
\title{Entanglement at the interplay between single- and many-bodyness}
\author{Jose Reslen}\email{josereslen@mail.uniatlantico.edu.co}
\affiliation{Coordinaci\'on de F\'{\i}sica, Universidad del Atl\'antico,
Carrera 30 N\'umero 8-49, Puerto Colombia.}%
%
\newcommand{\Keywords}[1]{\par\noindent{\small{\em Keywords\/}: #1}}
\begin{abstract} 
The tensor network representation of the ground state of a Bethe chain is
analytically obtained and studied in relation to its entanglement
distribution. Block entanglement displays a maximum at the interplay between
single- and many-bodyness. In systems of two fermions, tensor networks
describing ground states of interacting Hamiltonians cannot be written as a
sequence of next-neighbor unitaries applied on an uncorrelated state, but need
four-next-neighbor unitaries in addition. This differs from the idea that the
ground state can be obtained as a sequence of next-neighbor operations applied
on a tensor network. The work uncovers the transcendence of the notion of
many-bodyness in the implementation of protocols based on matrix product
states. 
\\
%
\end{abstract}
\maketitle
\section{Introduction}
A common strategy to deal with a complex problem is to divide it into small
parts that could be treated independently and with reduced difficulty.  This
is the essence of many methods based on Matrix Product States (MPS)
\cite{cirac}: A quantum state that spans over a wide space is described in
terms of a basis that makes it possible to operate at a local level, whether
it be to minimize the energy, as in Density Matrix Renormalization Group
(DMRG) \cite{DMRG} or to apply unitary transformations, as in Time Evolving
Block Decimation (TEBD) \cite{TEBD}, along with a long list of alternative
approaches \cite{Ulrich}. Despite the extended use of these methods and the
fact that they sometimes produce unsatisfactory results, there seems to be
hardly any questioning regarding their compatibility with the structure of the
states they intent to determine.  This might happen because of a lack of
examples in which the ground state of a many-body system could be obtained
exactly in MPS terms. The possibility of getting exact tensor networks
constitutes a powerful tool in the study of quantum systems.  On the one hand,
systems with exact tensorial representations serve as benchmarks for
simulation protocols based on MPS \cite{lewis}. On the other hand, such
representations can be used in combination with tensor network techniques, for
instance, as initial states in studies of quench dynamics \cite{quench}.
Typically, the suitability of MPS methods in one-dimensional systems is gauged
by the amount of entanglement between complementary blocks, which ultimately
defines the number of bond links of the tensor network. A celebrated result is
that block entanglement displays logarithmic growth as a function of the block
size at criticality, while saturates otherwise \cite{vidal2}, in which case it
is said that the state follows an area law \cite{area_law}. It then seems that
failure to follow an area law is given as the main cause why conventional
tensor network algorithms sometimes miscalculate the ground state. This being
so, it seems valid to ask why the problem cannot be solved simply by
increasing the bond dimension, after all, the fact that a method be
inefficient does not mean that same method be faulty.  Moreover, given the
importance that locality has on the formulation of tensor network methods, How
accurate is it assume that noncritical short-ranged Hamiltonians have local
ground states {\it at an operational level}? And finally, Why tensor network
methods perform much better on non-interacting- than on interacting-systems?

Much of the effort devoted to the development of MPS protocols has been
motivated by the desire to explore novel phases of matter, especially in
relation to the development of quantum entanglement. Considerable attention
has been awarded to systems that display established physical features such as
phase transitions \cite{amico} or chaos \cite{reslen_casati,lerose}. Such
features encompass a rise in the state's complexity that often leads to the
enhancement of entanglement. This study intents to characterize the
entanglement response as the ground state passes from a single-body- to a
many-body-profile, i.e., from being an eigenstate of the Hamiltonian's
single-body terms to being an eigenstate of the Hamiltonian's many-body terms.
This is done in an exactly solvable model consisting of a chain with two
interacting fermions. The analytical solution is recast in MPS description and
the resulting tensor network is scanned to obtain the entanglement.
Outstandingly, entanglement exhibits non-trivial behavior over the interlude
between single- and many-bodyness. Unexpectedly, the recasting procedure
reveals structure differences between single-body- and
many-body-representations, namely, while in the single-body case the state can
be expressed as a product of next-neighbor unitaries applied on an
uncorrelated state, in the many-body case the state requires
four-nearest-neighbor unitaries as well. This challenges the conception that
in the latter case the state can be obtained as a sequence of next-neighbor
operations, which is central to most MPS methodologies. 

A fermion chain, as originally proposed by Bethe \cite{bethe}, is described by
the Hamiltonian
\begin{gather}
\hat{H} =  \sum_{j=1}^N J (\hat{c}_j^\dagger \hat{c}_{j+1} +
\hat{c}_{j+1}^\dagger \hat{c}_j) + 
U \hat{c}_{j+1}^\dagger \hat{c}_{j+1} \hat{c}_{j}^\dagger \hat{c}_{j}.
\label{e07221}
\end{gather}
Ladder operators represent spinless fermion modes with standard
anticommutation identities $\{\hat{c}_j,\hat{c}_k^\dagger \} = \delta_j^k$ and
$\{\hat{c}_j,\hat{c}_k \} = 0$. Integer $N$ is the number of sites or
single-body states in the chain. Constants $J$ and $U$ modulate the intensity
of hopping and interaction, respectively. The system energy-scale is set by
making $J=1$, turning $U$ dimensionless.  Boundary conditions are periodic:
$\hat{c}_{j+N} = \hat{c}_{j}$. The total number of particles in the chain is
two.  Although no phase transition develops, the terms involved in the
Hamiltonian tend to induce different characters on the ground state, namely, a
single-body state when $U=0$ and a many-body state when $J=0$.  When $U$ is
positive fermions tend to repeal each other, but results ahead show no
substantial difference in the entanglement response between this and the
interactionless case. When $U$ is negative particles attract each other. It is
in this case that a truly many-body ground state gets to develop. In both
instances the model spectrum is exactly solvable by Bethe ansatz. The details
regarding the formulation of this ansatz in operator space can be consulted in
appendix \ref{a07261}. The preparation of Bethe eigenstates in a
quantum computer has recently been studied in \cite{economou}. The
entanglement distribution of half-filled chains described by (\ref{e07221})
has been studied in \cite{casiano1,casiano2,casiano_tesis} in connection to
entanglement usability. 
\section{MPS implementation of a two-fermion state}
\label{s2208141}
Since block entanglement is most easily obtained from a MPS representation, we
now discuss how to get this representation from a state originaly written in a
Fock basis.
In appendix \ref{a07261} it is shown that an eigenstate of Hamiltonian
(\ref{e07221}) with two fermions can be written as
\begin{gather}
|E \rangle = \sum_{m_1=1}^{N-1} \sum_{m_2=m_1+1}^{N} a_{m_1 m_2}
\hat{c}_{m_1}^\dagger \hat{c}_{m_2}^\dagger |0 \rangle,
\label{e07161}
\end{gather}
where
\begin{gather}
a_{m_1 m_2} = q_1 e^{i k_1 m_1 + i k_2 m_2} +  q_2 e^{i k_2 m_1 + i k_1 m_2},
\label{e01071}
\end{gather}
as long as $m_1<m_2$. Otherwise $a_{m_1 m_2} = 0$. Non-degenerate eigenstates
must have real coefficients because Hamiltonian (\ref{e07221}) is represented
by a real symmetric matrix in a Fock basis. In order to ensure real
coefficients, only ground states of chains with odd $N$ are considered.

Equation (\ref{e07161}) can also be written as
\begin{gather}
|E \rangle = \frac{1}{2} \sum_{m_1=1}^{N} \sum_{m_2=1}^{N} A_{m_1 m_2} \hat{c}_{m_1}^\dagger \hat{c}_{m_2}^\dagger |0 \rangle,
\label{e07182}
\end{gather}
so that
\begin{gather}
A_{m_1 m_2} = a_{m_1 m_2} - a_{m_2 m_1}.
\label{e07271}
\end{gather}
It becomes in this way noticeable that the coefficients form a {\it real
antisymmetric matrix} $A_{m_1 m_2} = -A_{m_2 m_1}$. According to spectral
theory \cite{youla}, matrix (\ref{e07271}) can be factorized as 
\begin{gather}
\hat{A} = \hat{U} \hat{\Lambda} \hat{U}^T,
\label{e07181}
\end{gather}
where $\hat{U}$ is orthogonal (real unitary) and 
\begin{gather}
\hat{\Lambda} = 
\left (
\begin{array}{ccccc}
0 & \alpha_1 & 0 & 0 & \hdots  \\
-\alpha_1 & 0 & 0 & 0 & \hdots  \\
0 & 0 &  0 & \alpha_2 & \hdots  \\
0 & 0 & -\alpha_2  & 0 & \hdots  \\
\vdots & \vdots & \vdots  & \vdots & \ddots   
\end{array}
\right ). 
\end{gather}
The $\alpha$'s are positive coefficients. When the matrix dimension is odd
there is at least one column and row of zeros. The state can be written as
\begin{gather}
|E \rangle = \sum_{j} \alpha_j \hat{f}_{2j-1}^\dagger \hat{f}_{2j}^\dagger  |0 \rangle,
\label{e07291}
\end{gather}
where the $\hat{f}^\dagger$s are genuine fermionic modes related to the
original ones by
\begin{gather}
\left [
\begin{array}{c}
\tilde{f}_1^\dagger    \\
\tilde{f}_2^\dagger    \\
\vdots         \\
\tilde{f}_{N}^\dagger
\end{array}
\right ] =
\left [
\begin{array}{cccc}
U_{1,1} & U_{2,1}  & \dots   & U_{N,1} \\
U_{1,2} & U_{2,2}  & \dots  & U_{N,2} \\
\vdots    & \vdots              &  \vdots &  \vdots \\
U_{1,N}& U_{2,N} & \dots  & U_{N,N}
\end{array}
\right ]
\left [
\begin{array}{c}
\tilde{c}_{1}^\dagger    \\
\tilde{c}_{2}^\dagger    \\
\vdots      \\
\tilde{c}_{N}^\dagger
\end{array}
\right ].
\label{star}
\end{gather}
The elements $U_{jk}$ above are the coefficients of matrix $\hat{U}$ in
(\ref{e07181}), although the arrangement in this expression corresponds to
$\hat{U}^T$. The process of decomposing, or folding, a relation like
(\ref{star}) in similar scenarios has been described in references
\cite{reslen5,reslen6} for non-interacting fermions and
\cite{ReslenRMF,reslen4} for bosons. The scheme is based on the observation
that an unitary transformation like (\ref{e07292}) on state (\ref{e07291}) has
effects only on the columns of matrix (\ref{star}) corresponding to the modes
involved in said transformation. In this way, let us consider an unitary
operation involving neighbour modes thus
\begin{gather}
\hat{W}_{j,1} = e^{-i\theta_{j,1} \hat{h}_j }, \text{ } \hat{h}_j = \frac{1}{2 i}
(\hat{c}_{j+1}^\dagger \hat{c}_j - \hat{c}_j^\dagger \hat{c}_{j+1}).
\label{e07292}
\end{gather}
The effect of this transformation on the coefficients of the first row of
matrix (\ref{star}) can be determined by noticing how it operates on a sum of
neighbor modes
\begin{gather}
\hat{W}_{1,j} \left (  U_{j+1,1} \hat{c}_{j+1}^{\dagger} +  U_{j,1}
\hat{c}_{j}^{\dagger} \right ) \hat{W}_{1,j}^{-1}    \nonumber \\
= \left ( U_{j+1,1} \cos \left( \frac{\theta_{j,1}}{2} \right) - U_{j,1} \sin \left( \frac{\theta_{j,1}}{2} \right)  \right )\hat{c}_{j+1}^{\dagger} \nonumber \\
+ \left ( U_{j+1,1} \sin \left( \frac{\theta_{j,1}}{2} \right) + U_{j,1} \cos \left( \frac{\theta_{j,1}}{2} \right)  \right )\hat{c}_{j}^{\dagger}.
\label{eq:9}
\end{gather}

As a result, the contribution of $\hat{c}_{j+1}^{\dagger}$ can always be
suppressed by choosing the appropriate angle, namely,

\begin{gather}
\tan \left( \frac{\theta_{j,1}}{2} \right) = \frac{U_{j+1,1}}{U_{j,1}}.
\label{eq:11}
\end{gather}
The transformation can initially be applied on the last two columns, i.e., the
last two modes, leaving an updated matrix in (\ref{star}) with the following
shape
\begin{gather}
\left [
\begin{array}{cccc}
U_{1,1}   & \dots & U_{N-1,1}' & 0 \\
U_{1,2}    & \dots & U_{N-1,2}'  & U_{N,2}' \\
\vdots              & \vdots    &  \vdots &  \vdots \\
U_{1,N}   & \dots & U_{N-1,N}' & U_{N,N}'
\end{array}
\right ].
\end{gather}
Analogous transformations can be applied in order to eliminate the
coefficients on the first row, except the element on the top-left corner,
which cannot be eliminated in the same fashion. The matrix then takes a form
along the lines of 
\begin{gather}
\left [
\begin{array}{cccc}
U_{1,1} & 0  & \dots   & 0 \\
0 & U_{2,2}  & \dots  & U_{N,2} \\
\vdots    & \vdots              &  \vdots &  \vdots \\
0 & U_{2,N} & \dots  & U_{N,N}
\end{array}
\right ].
\end{gather}
Primes have been dropped to facilitate the reading but notice that
non-vanishing elements in this matrix are in general different from the
elements that appear in (\ref{star}). All the coefficients below $U_{1,1}$ in
the first column must vanish because that is the only way how anticommutation
relations among the updated modes are conserved. A similar protocol is then
applied on the second row: The coefficients are canceled using next-neighbor
unitary transformations. The process however must not involve the first mode
because this would unfold the first row. This cancellation mechanism is
repeated on the other columns until the matrix is completely diagonal. This
reduction is reflected on the state in such a way that at the end it displays
a simpler structure, specifically
\begin{gather}
|E \rangle \rightarrow |E_R \rangle = \sum_{j} \alpha_j \hat{c}_{2j-1}^\dagger
\hat{c}_{2j}^\dagger  |0 \rangle.
\label{e07191}
\end{gather}
In order to get the original state (\ref{e07182}) in MPS representation, the
first step is to obtain the tensor description of state (\ref{e07191}).  Next,
the inverses of the folding transformations outlined above must be applied on
this state, taking care of following a reverse order with respect to the
sequence used to go from the original state to the reduced one.  We call this
the unfolding. Formally, this conversion protocol can be represented through
the following identity
\begin{gather}
|E \rangle =
\prod_{k=N-1}^1  \prod_{j=k}^{N-1} \hat{W}_{j,k}^{-1} |E_R \rangle.
\label{e2208144}
\end{gather}
As all the transformations involved in the unfolding are next-neighbor and
unitary, their effect on MPS networks can be established by means of the
update protocols of reference \cite{vidal}. Details of the application of this
protocols to a fermion chain can be found on the first appendix of reference
\cite{reslen5}. In order to complete the transformation, state (\ref{e07191})
must be written as a tensor network. For this purpose let us consider the
state in its explicit form
\begin{gather}
|E_R \rangle = \alpha_1 |1100...00 \rangle + \alpha_2 |0011...00 \rangle 
\nonumber \\
+ ... +  \alpha_j |...001100... \rangle + ...
\label{e2208041}
\end{gather}
\begin{figure}
\begin{center}
\includegraphics[width=0.45\textwidth,angle=0]{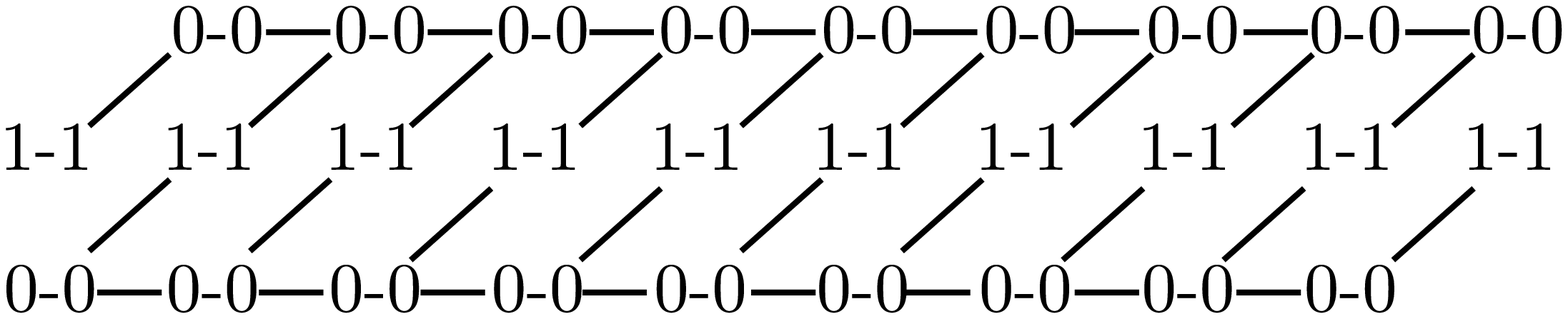}
\caption{Schematics of tensor connections of state (\ref{e2208041}).}
\label{fig1}
\end{center}
\end{figure}
The elements of a canonical MPS representation on a chain can be seen as the
coefficients of an expansion using local states plus Schmidt vectors as a
basis, like so
\begin{gather}
| E_R \rangle = \sum_{\mu} \sum_{\nu}  \sum_{k=0}^1 \lambda_{\mu}^{L-1} 
\Gamma_{\mu\nu}^{k L} \lambda_{\nu}^{L} |\mu \rangle |k \rangle |\nu \rangle.
\label{e08031}
\end{gather}
Vector $|k\rangle$ represents the local Fock state at site $L$. Kets $|\mu
\rangle$ and $|\nu \rangle$ are Schmidt vectors spanning between the chain's
edges and the sites to the left and right of site $L$, respectively. As
Schmidt vectors they must satisfy $\langle \mu' | \mu \rangle = 0$ and
$\langle \nu' | \nu \rangle = 0$. Real numbers $\lambda_\mu^{L-1}$ and
$\lambda_\nu^L$ are the Schmidt coefficients associated to $|\mu \rangle$ and
$|\nu \rangle$, respectively. Tensor $\Gamma_{\mu \nu}^{k L}$ contains the
coefficients of the superposition. There is a set of coefficients
$\lambda_\mu^{L-1}$, $\lambda_\nu^L$ and $\Gamma_{\mu \nu}^{k L}$ for each $L$
and the collection of these sets over $L=1,2,...,N$ form a canonical MPS
representation of $| E_R \rangle$.  Normally, the canonical decomposition of a
quantum state is nontrivial, but for a state like (\ref{e2208041}) the
distribution of Schmidt vectors can be discerned through the connection map
shown in figure \ref{fig1}. Each line is a $\lambda$ and each number is a
$\Gamma$. For instance, it can be seen that the coefficients associated to
$L=1$ on the chain's left end can be chosen as
\begin{gather}
\lambda_1^1=\alpha_1 \text{, } \lambda_2^1=1 \text{, }
\Gamma_{1 1}^{1 1}=\Gamma_{1 2}^{0 1}=1.
\label{e08042}
\end{gather}
A similar derivation can be made for the coefficients of the second and third
site. The coefficients on the bulk of the chain in figure \ref{fig1} display a
periodic pattern, being
\begin{gather}
\lambda_1^L=\lambda_2^L=\lambda_3^L=\lambda_4^L=1, \nonumber \\
\Gamma_{2 2}^{1 L} = \Gamma_{1 1}^{0 L} = \Gamma_{3 3}^{0 L} = \Gamma_{3 4}^{0 L} = 1,
\label{e08043}
\end{gather}
for $L$ even and
\begin{gather}
\lambda_1^L=\lambda_3^L=1 \text{, } \lambda_2^L=\alpha_{\frac{L+1}{2}}, \nonumber \\
\Gamma_{3 2}^{1 L} = \Gamma_{1 1}^{0 L} = \Gamma_{2 1}^{0 L} = \Gamma_{4 3}^{0 L} = 1,
\label{e08044}
\end{gather}
for $L$ odd. Once the tensor network describing $|E\rangle$ is obtained
applying next-neighbor unitaries to the MPS representation of $|E_R\rangle$,
the entanglement structure can be studied using the entanglement measures
referenced below. 
\section{Entanglement measures}

\subsection{Block entropy}
Given a pure state of a quantum chain divided in two continuous blocks, the
entanglement between these blocks can be measured by the von Neumman entropy
\cite{area_law, plenio}:
\begin{gather}
S_L = -Tr(\hat \rho_L \log \hat \rho_L),
\end{gather}
where $\hat \rho_L$ is the reduced density matrix of the block spanning from
the first- to the $L$th-site: 
\begin{gather}
\hat \rho_L = Tr_{[L+1:N]}(| E \rangle \langle E |).
\end{gather}
The bracket symbol above represents the tracing out of all the degrees of
freedom between sites $L+1$ and $N$.  Vanishing values of $S_L$ indicate the
state is separable with respect to the blocks involved, i.e., it
can be written as a product state of such blocks:
\begin{gather}
|E\rangle = |\psi_{[1:L]}\rangle |\psi_{[L+1:N]}\rangle.
\end{gather}
Moreover, von Neumman entropy is invariant under local operations performed on
either of the considered blocks, which makes it a perfect quantifier of
bipartite entanglement.  In addition to its relevance as a quantum resource,
block entanglement is associated with the simulation cost incurred in
computing the state using MPS methods since $S_L$ quantifies the level of
locality (statistical independence with respect to the rest of the chain) of
$\hat {\rho}_L$. If the state is given as a tensor network in canonical form,
the block entropy can be readily calculated from the coefficients as follows
\begin{gather}
S_L = -\sum_{k} \lambda_k^L \log \lambda_k^L.
\end{gather}
\subsection{Two-body entropy (Many-bodyness)}
In a system with two fermions it is possible to establish a measure of
many-bodyness in close parallel to von Neumman entropy. In fact, just as $S_L$
is invariant under local operations, a many-bodyness measure should display
invariance under single-body unitary-operations. From equation (\ref{e07291})
it can be seen that such operations have the following effect on the state
\begin{gather}
\hat V |E \rangle = \sum_{j} \alpha_j \overbrace{\hat V \hat{f}_{2j-1}^\dagger
\hat V^{-1}}^{\hat{f'}_{2j-1}^\dagger} \overbrace{\hat V \hat{f}_{2j}^\dagger
\hat V^{-1}}^{\hat{f'}_{2j}^\dagger} |0 \rangle.
\end{gather}
Primed operators represent standard fermionic modes, owing to the unitarity of
$\hat V$. As a result, the transformed state maintains the structure of
equation (\ref{e07291}), showing coefficients $\alpha_j$ are invariant under
single-body unitary-operations. On these grounds, the following criterion is
proposed
\begin{gather}
S = -\sum_{j} \alpha_j \log \alpha_j.
\label{e2208231}
\end{gather}
The limit case $S=0$ corresponds to a situation where the state is simply
given by
\begin{gather}
| E \rangle = \hat{f}_1^\dagger \hat{f}_{2}^\dagger|0 \rangle,
\end{gather}
which genuinely describes an eigenstate of free fermions. Therefore, only
states with authentic many-body correlations can display nonvanishing values
of $S$.  The coefficients $\alpha$ can be obtained from the decomposition of
matrix $A$ shown in (\ref{e07181}) and as such are part of the protocol
employed to get the MPS representation of $|E\rangle$ introduced above.
\section{Results}
\begin{figure}
\begin{center}
\includegraphics[width=0.33\textwidth,angle=-90]{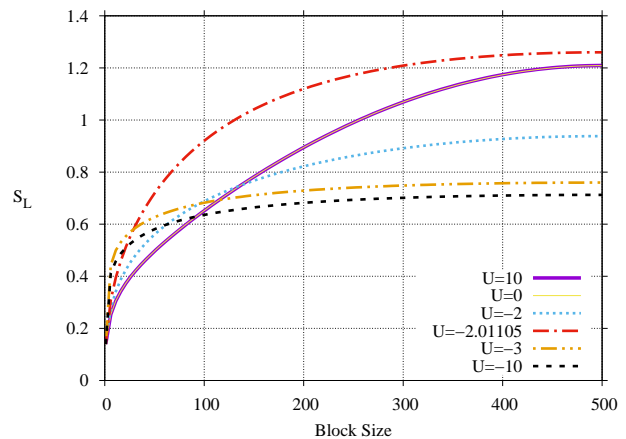}
\caption{Block entropy vs block size in the ground state of Hamiltonian
(\ref{e07221}) with two fermions and $N=1001$. Curves for $U=0$ and $U=10$
appear superimposed, showing there is little change in the system's response
over the whole range of positive values of $U$. Block entropy saturates on
every case, but a chain with maximum entanglement is observed for a negative
value of $U$, indicating that a local interaction can have an enhancing effect
on long-range entanglement when properly combined with hopping, even in the
absence of phase transitions. In all cases shown $J=1$.
}
\label{fig2}
\end{center}
\end{figure}
\begin{figure}
\begin{center}
\includegraphics[width=0.33\textwidth,angle=-90]{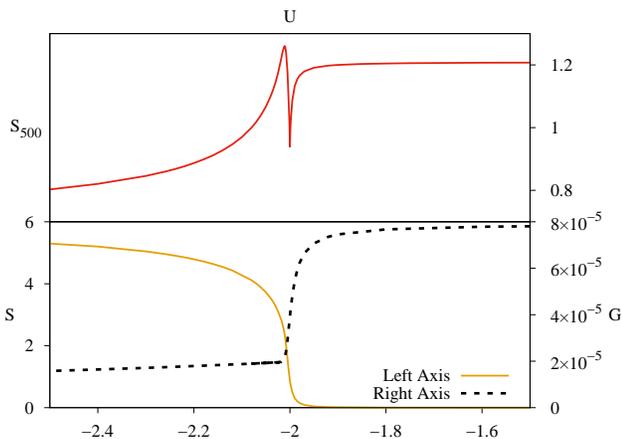}
\caption{Top: Entanglement between the two halves of a chain describing the
ground state of Hamiltonian (\ref{e07221}) with two fermions and $N=1001$.  A
local minimum is observed at $U=-2$ and a global maximum at around $U=-2.011$.
Bottom, left axis: Two-body entropy on the same system. Right axis: Energy
gap. A sharp fluctuation in the block entanglement takes place around a short
window of values of $U$ where the two-body entropy is halfway between zero and
its saturation value, clearly indicating a correlation involving entanglement
and the interplay between single- and many-bodyness. In all cases shown $J=1$.
}
\label{fig3}
\end{center}
\end{figure}
Block entanglement as a function of block size is depicted in figure
\ref{fig2} for a Bethe chain with two fermions and $N=1001$. The dependency
pattern is basically the same over the values of $U$ where two-body entropy,
which can be seen in the bottom panel of figure \ref{fig3}, is small relative
to its saturation value. Specifically, this occurs over the range of values of
$U/J$ between $-1.9$ and infinity. Just under $-1.9$, the maximum value of
block entropy displays a sharp fluctuation, as can be seen in the top panel of
figure \ref{fig3}. A global maximum and a local minimum show up over a short
window in the interaction domain. Furthermore, the gap remains finite
throughout the parameter range, showing no phase transition, either
conventional or topological, takes place. However, block entanglement displays
a maximum at a point where the value of $S$ shows a clear contribution, yet
not an overtaking, of many-bodyness. This suggests that it is not
many-bodyness alone that causes entanglement to peak, but rather the interplay
between single- and many-bodyness, since entanglement fluctuates at finite
values of $U$ where neither interaction nor hopping is dominant. From a
technical perspective, the interaction term alone has eigenstates that span
over the local sector of the Fock basis, while the hopping term alone displays
eigenstates that span on the single-body sector of the spectrum. It is
therefore only when the Hamiltonian parameters stand in a range where both
interaction and hopping prevail that the superposition distribution spreads
over a wider sector of the Hilbert space and entanglement develops. The fact
that this can happen even in the absence of a phase transition suggests that
the entanglement that spontaneously arises in physical systems is more
dependent on the level of access to the elements of a basis than it might be
on criticality itself.  In this sense, the half-filled chain has the maximum
entanglement potential. 
\section{Discussion}
Interestingly, there is another way of writing $|E_R\rangle$ in
(\ref{e2208041}) as a MPS. For this, let us see that over this state one can
apply the next unitary transformation
\begin{gather}
\hat M_1 = e^{i \phi_1 \hat m_1} \text{, } \hat m_1 = \frac{1}{2 i} (|\omega_2
\rangle \langle \omega_1 | - |\omega_1 \rangle \langle \omega_2 |),
\end{gather}
where
\begin{gather}
| \omega_1 \rangle = |1100...00\rangle \text{, } | \omega_2 \rangle =
|001100...00\rangle. 
\label{e2208141}
\end{gather}
The result being
\begin{gather}
\hat M_1 |E_R \rangle = \left( \alpha_1 \cos \frac{\phi_1}{2} - \alpha_2 \sin
\frac{\phi_1}{2} \right ) | \omega_1 \rangle \nonumber \\
+ \left( \alpha_2 \cos \frac{\phi_1}{2} + \alpha_1 \sin \frac{\phi_1}{2} \right ) | \omega_2 \rangle + \alpha_3 |\omega_3 \rangle ...,
\end{gather}
with $|\omega_3 \rangle$ having a meaning equivalent to that of $|\omega_1
\rangle$ and $|\omega_2 \rangle$ in (\ref{e2208141}). The coefficient of
$|\omega_1 \rangle$ can always be made to vanish by choosing: 
\begin{gather}
\tan \frac{\phi_1}{2} = \frac{\alpha_2}{\alpha_1}.
\label{e2208142}
\end{gather}
In similarity to the folding protocol described in section \ref{s2208141}, an
analogous transformation can be applied to this reduced state, this time
involving $|\omega_2 \rangle$ and $|\omega_3 \rangle$ in order to eliminate
$|\omega_2 \rangle$. The process continues until one last Fock state remains.
The original state can then be obtained by reversing the procedure:
\begin{gather}
|E_R \rangle = \prod_l \hat {M}_l^{-1} |00...011\rangle.
\label{e2208143}
\end{gather}
Replacing this expression in (\ref{e2208144}) it is possible to express
$|E\rangle$ as a series of unitary transformations applied on a simple Fock
state easily expressible as MPS. In principle, this possibility offers another
simulation path implementable by MPS. However, on closer inspection a
potential complication arises: Transformations of the type $\hat {M}_l^{-1}$
in (\ref{e2208143}) are not like the $\hat{W}_{j,k}^{-1}$ in (\ref{e2208144}).
The latter operate on next neighbors, while the former operate on sets of four
nearest neighbors. This contrast with the fact that current MPS protocols are
based on the repeated application of next-neighbor operations on tensor
networks. This does not mean that the $\hat{M}_l^{-1}$s cannot be implemented
over a tensor network, but such an implementation requires additional
considerations. As a solution, one could think of establishing a local space
of the original chain consisting not of one site but of two sites, so that a
transformation involving four-nearest neighbors could be equivalently realized
as a next-neighbor transformation. Contrariwise, when interaction is zero
there is no need to consider four-nearest neighbor transformations because in
that case two-body entropy vanishes and $|E_R\rangle$ is a Fock state, meaning
that the decomposition depends entirely on genuine next-neighbor unitaries.
This explains why methods based on MPS work so well on non-interacting
systems. It also suggest a way of implementing variational MPS to find ground
states of interacting systems: Instead of initializing the network as an
uncorrelated state, the initial configuration should include many-body
correlations in analogy to $|E_R\rangle$ in (\ref{e2208041}). In this way, the
coefficients of this initial configuration become variational parameters, just
as the elements of the tensor network, and should be determined as part of an
energy-minimization algorithm. Another way of using tensor networks to find
ground states is through imaginary-time evolution \cite{orus}. This requires
to split the evolution operator into a product of unitaries using a
Suzuki-Trotter expansion \cite{suzuki}.  The approximation parameter is the
time slice. The smaller the time slice the better the expansion accuracy.
However, if the ground state does contain many-body correlations the following
contradiction comes into play: Using small time slices is good to preserve the
canonical representation and keep errors under control, but it is bad to
approach the authentic ground state because such a state cannot be written in
terms of a product of next-neighbor unitaries on an uncorrelated initial
state. This is consistent with observations that the simulation accuracy peaks
at relatively high values of the time slice and decreases as the time slice is
shrunk \cite{reslen4}.  The solution is then to incorporate many-body
correlations on the initial state, although this add parameters that must be
determined variationally, as before. This shows how important it is to
establish a measurement of many-bodyness in systems with more than two
fermions.  Unfortunately, equation (\ref{e2208231}) is not scalable because
there is no way of factorizing a tensor of more than two indices in a way
analogous to (\ref{e07181}), which is related to the fact that there is no way
of effectuating higher order Schmidt decompositions \cite{peres}. Even so,
there might be other many-body criteria with equivalent functionality.  Ground
states displaying area laws do not necessarily have small many-body
correlations. From figures \ref{fig2} and \ref{fig3} it can be seen that
systems with very negative $U$ show rapid entanglement-saturation, but also
high values of $S$. In spite of following area laws, these systems are most
likely to present convergence issues when simulated via next-neighbor
MPS-methods if many-body correlations are not incorporated on the initial
state. This is the opposite of what happens for positive values of $U$, where
the value of many-bodyness is notoriously marginal. Such a range circumscribes
a set of highly interacting systems whose ground states can be effectively
simulated using conventional tensor networks techniques.
\section{Conclusions}
The MPS structure of a two-fermion chain exactly solvable by Bethe ansatz has
been analytically obtained and used to study the relation between entanglement
and many-bodyness. Maximally entangled chains are observed when the state
stands halfway between single- and many-bodyness. The study shows how to
obtain any eigenstate as a product of unitary operators acting on an
uncorrelated state. This decomposition reveals that states without many-body
correlations can be written as a product of next-nearest-neighbor unitaries,
whilst states with many-body correlations need four-nearest-neighbor unitaries
in addition. This feature clashes with the assumption that the ground state of
interacting Hamiltonians can be obtained as a sequence of next-neighbor
operations, a premise that is fundamental in the formulation of many MPS
methods. A potential solution is to incorporate many-body correlations on the
initial state, but the quantification of such correlations in systems with
more than two fermions needs characterization.  Ultimately, block entanglement
is not the only determining factor in MPS simulations, but also the amount of
many-body correlations contained by the target state. Conventional MPS methods
are better suited to systems where the ground state has little many-body
entropy.
%
%
%
%
%
%

%
%
%
%
%
%
\appendix
\section{Bethe Anzats}
\label{a07261}
Let us consider a chain where two fermions can tunnel between adjacent places
and interact when they simultaneously occupy neighboring sites of a chain with
$N$ sites \cite{bethe}. The system's physics is described by the Hamiltonian
\begin{gather}
\hat{H} = \overbrace{J \sum_{j=1}^N (\hat{c}_j^\dagger \hat{c}_{j+1} +
\hat{c}_{j+1}^\dagger \hat{c}_j)}^{\hat{H}_J} + \overbrace{U \sum_{j=1}^N
\hat{c}_{j+1}^\dagger \hat{c}_{j+1} \hat{c}_{j}^\dagger}^{\hat{H}_U} \hat{c}_{j}.
\label{e08041}
\end{gather}
Mode operators satisfy standard fermionic rules $\{\hat{c}_j,\hat{c}_k^\dagger
\} = \delta_j^k$ and $\{\hat{c}_j,\hat{c}_k \} = 0$. Constants $J$ and $U$
determine the intensity of hopping and interaction respectively. The chain
displays periodic boundary conditions, so that $\hat{c}_{j+N} = \hat{c}_{j}$.
An eigenstate of a chain is hypothesized to be of the form
\begin{gather}
|E \rangle = [q_1 \hat{\psi}(k_1,k_2) + q_2 \hat{\psi}(k_2,k_1)] |0\rangle,
\label{e05281}
\end{gather}
where
\begin{gather}
\hat{\psi}(k_1,k_2) = \sum_{m_1=1}^{N-1} \sum_{m_2=m_1+1}^{N} e^{i k_1 m_1 + i
k_2 m_2} \hat{c}_{m_1}^\dagger \hat{c}_{m_2}^\dagger.
\label{e05311}
\end{gather}
Coefficients $q_1$, $q_2$, $k_1$ and $k_2$ are in general complex and must be
adjusted to make $|E\rangle$ and eigenstate of $\hat{H}$. Ket $|0\rangle$
represents a state without fermions. In order to test the proposed solution,
the effect of the hopping term is calculated, so obtaining the following
expression
\begin{gather}
|\hat{H}_J E \rangle /J = 2 (\cos k_1 + \cos k_2) |E \rangle \nonumber \\
- (q_1 + q_2)
(1 + e^{i(k_1+k_2)}) \sum_{m=1}^N e^{i(k_1+k_2)m} \hat{c}_m^{\dagger}
\hat{c}_{m+1}^{\dagger} |0\rangle   \nonumber \\
-(q_1 + q_2 e^{i k_1 N}) \sum_{m=1}^{N-1} e^{i k_2 m} ( \hat{c}_1^\dagger
\hat{c}_m^\dagger +  e^{i k_1} \hat{c}_m^\dagger \hat{c}_N^\dagger)|0\rangle
 \nonumber \\
-(q_2 + q_1 e^{i k_2 N}) \sum_{m=1}^{N-1} e^{i k_1 m} ( \hat{c}_1^\dagger
\hat{c}_m^\dagger +  e^{i k_2} \hat{c}_m^\dagger \hat{c}_N^\dagger)|0\rangle.
\label{e07252}
\end{gather}
When $k_1 \ne k_2$ terms in the last two lines can be canceled by setting
\begin{gather}
q_2 = - q_1 e^{i k_2 N} = - q_1 e^{-i k_1 N} \label{e05172} \\
\rightarrow e^{i (k_1+k_2) N} = 1. 
\label{e05171}
\end{gather}
The latter requirement can be met through the following relation
\begin{gather}
k_1+k_2 = \frac{2 \pi n}{N}, \text{ } n = 0,1,...,N-1.
\label{e05233}
\end{gather}
Regarding the interaction part of the Hamiltonian, it can be shown that 
\begin{gather}
|\hat{H}_U E \rangle/U = (q_1 e^{i k_2} + q_2 e^{i k_1}) \sum_{m=1}^N
e^{i(k_1+k_2) m} \hat{c}_m^\dagger \hat{c}_{m+1}^\dagger |0\rangle.
\label{e05231}
\end{gather}
Both (\ref{e05172}) and (\ref{e05171}) have been utilized in the calculation
leading to (\ref{e05231}). Using the expressions obtained above for hopping 
and interaction the eigenvalue equation can be formulated as
\begin{gather}
|\hat{H} E \rangle = 2 J (\cos k_1 + \cos k_2) |E\rangle \nonumber \\
+ \left( U (q_1 e^{i k_2}
+ q_2 e^{i k_1}) - J(q_1 + q_2)(1 + e^{i(k_1 + k_2)}) \right) \nonumber \\
\times \sum_{m=1}^N e^{i(k_1+k_2) m} \hat{c}_m^\dagger \hat{c}_{m+1}^\dagger |0\rangle.
\label{e06082}
\end{gather}
Using again equations (\ref{e05172}) and (\ref{e05171}) it can be shown that
the term in parentheses vanishes for values of $z=e^{-i k_1}$ being solutions
of 
\begin{gather}
(1+\alpha_n) z^N - \gamma z^{N-1} + \gamma \alpha_n  z - (1+\alpha_n) = 0,
\label{e05271}
\end{gather}
where
\begin{gather}
\alpha_n = e^{i \frac{2 \pi n}{N}}, \text{ } n = 0,1,...,N-1, \text{ and } \gamma =
\frac{U}{J}, \text{ } J\ne0.
\label{e05312}
\end{gather}
Notice that if $\alpha_n = -1$, which can happen only when $N$ is even, the
order of polynomial (\ref{e05271}) is reduced.  Once $k_1$ has been determined
from $z$, $k_2$ can be found from (\ref{e05233}).  These constants can be used
to find the system energy from
\begin{gather}
E = 2 J (\cos k_1 + \cos k_2).
\label{e06041}
\end{gather}
Constants $q_1$ and $q_2$ can be determined using equation (\ref{e05172}) and
the normalization condition for eigenstates. From the whole set of solutions
that can be built in this way, some might be redundant. This happens because
swapping $k_1 \leftrightarrow k_2$ does not change state (\ref{e05281}), as
can be implied from the next identity (derived using (\ref{e05172}) and
(\ref{e05171})) 
\begin{gather}
q_1 \hat{\psi}(k_2,k_1) + q_2 \hat{\psi}(k_1,k_2) \nonumber \\
= -e^{i k_1 N } (q_1\hat{\psi}(k_1,k_2) + q_2 \hat{\psi}(k_2,k_1)).
\end{gather}
Hence, one can discard a pair $(k_1',k_2')$ whose relation to another
complying pair $(k_1,k_2)$ be $(k_1',k_2') = (k_2,k_1)$. It can happen that
some solutions provided by (\ref{e05271}) derive in complex $k$s. Such
solutions are valid.  Moreover, in this case $k_1=k_2^*$ in order for
relations (\ref{e05171}) and (\ref{e06041}) to hold. The following particular
cases must be  analysed separately
\begin{enumerate}
\item $\alpha_n = -1$ and $\gamma \ne 0$
For this $n = \frac{N}{2}$ and therefore $N$ must be even. Equation
(\ref{e05271}) is then reduced to
\begin{gather}
z^{N-2} + 1 = 0.
\label{e06021}
\end{gather}
Technically, $z=0$ is also a root but it does not lead to any meaningful
solution.  From (\ref{e06021}) it follows
\begin{gather}
z = e^{i \pi (2 j + 1)/(N-2)}, \text{ } j = 0,1,...,N-3.
\end{gather}
The rest of parameters can be found from $z$ according to the procedure
prescribed above.
\item $\gamma = 0$ $(U=0, J\ne0)$. 
An inspection of the solutions provided by equation (\ref{e05271}) in this
regime evidences a number of inconsistencies, in particular, some solutions
are not independent. In this case it is better to build independet solutions
using a different protocol. This can be done without major complications by
observing that in absence of interaction a solution can be written as
\begin{gather}
|E \rangle  = \sum_{m_1=1}^{N} e^{i k_1 m_1} \hat{c}_{m_1}^\dagger \sum_{m_2=1}^{N}
e^{i k_2 m_2} \hat{c}_{m_2}^\dagger |0\rangle.
\end{gather}
This form is a particular case of the Bethe solution in (\ref{e05281}) when
$q_2=-q_1$.  In order to guarantee orthogonal solutions, $k_1$ and $k_2$ must
take the next form
\begin{gather}
k_1 = 2 \pi j_1 / N, \text{ } k_2 = 2 \pi j_2 / N.
\end{gather}
Both $j_1$ and $j_2$ are integers that can take values between 1 and $N$ under
the constrain $j_1<j_2$. The corresponding energy can be obtained through
equation (\ref{e06041}) replacing $k_1$ and $k_2$. The solutions obtained in
this way are independent and form a complete set.
\item It happens that after solving equation (\ref{e05271}) and finding the
constants of interest, some solutions with $k_1 = k_2 = k$ turn up. The
significance of such solutions must be inspected. In this case the procedure
followed to arrive at (\ref{e05271}) is compromised because the terms in the
last two lines of equation (\ref{e07252}) group in a different way, namely
\begin{gather}
(q_1 + q_2)(1+ e^{i k N}) \sum_{m=1}^{N-1} e^{i k m} ( \hat{c}_1^\dagger
\hat{c}_m^\dagger +  e^{i k} \hat{c}_m^\dagger \hat{c}_N^\dagger)|0\rangle.
\label{e06083}
\end{gather}
In principle, it appears there are two ways of cancelling this contribution.
In the first place, one can set
\begin{gather}
k = \frac{\pi (2 j +1 )}{N}, \text{ } j = 0,1,...
\label{e06081}
\end{gather}
This is the same estimation of $k$ that can be obtained through (\ref{e05233})
by setting $k_1 = k_2 = k$ and selecting $n$ odd. However, relation
(\ref{e05172}) is no longer required to complete the cancellation. Since such
a relation has been used to get (\ref{e05271}), expression (\ref{e05271})
itself losses validity and any solution provided by it that might display
equal $k$s is unreliable. One can however group terms and prove that the term
in parenthesis in equation (\ref{e06082}) takes the next form
\begin{gather}
(q_1 + q_2)( U e^{i k} - J(1 + e^{2 i k})) = 0.
\end{gather}
Because at this stage $k$ is already given by (\ref{e06081}), the only
possibility is to make $q_2 = -q_1$, which is in contradiction with equation
(\ref{e05172}) for the values of $k$ taken in (\ref{e06081}), making it clear
that in this case equation (\ref{e05271}) is flawed. Hence it follows making
$q_1 = -q_2$ is the only way the whole term (\ref{e06083}) can be canceled.
Just as in the previous case, equation (\ref{e05172}) is not enforced and
therefore the solutions provided by (\ref{e05271}) cannot be admitted.
Constant $k$ is no longer conditioned by (\ref{e06081}).  However, making $k_1
= k_2$ and $q_2 = -q_1$ at the same time identically nullifies the original
anzats in (\ref{e05281}). Thus, there cannot be relevant solutions with equal
$k$s and therefore any solution arising from the protocol displaying such a
feature is discarded.
\item There is one solution that eludes the anzats. Let us consider an
eigenstate with the following structure
\begin{gather}
| E \rangle = q \sum_{m=1}^{N} e^{i k m} \hat{c}_{m}^\dagger
\hat{c}_{m+1}^\dagger|0 \rangle.
\label{e06101}
\end{gather}
Constant $q$ is essentially a normalization parameter. The effect of the
hopping term on this state can be written as
\begin{gather}
| \hat{H}_J E \rangle/q J = (1 + e^{i k}) \sum_{m=1}^{N} e^{i k j }
\hat{c}_j^\dagger \hat{c}_{j+2}^\dagger |0\rangle \nonumber \\
+ (1 - e^{i k N}) \hat{c}_{N}^\dagger \hat{c}_{2}^\dagger |0\rangle.
\end{gather}
The whole term can be nullified by making $k=\pi$, but only if $N$ is even.
Since state  (\ref{e06101}) is in general an eigenstate of $\hat{H}_U$, in
this specific case it becomes an eigenstate of $\hat{H}$ too. Constant $J$
makes no contribution to energy and the corresponding eigenvalue equation
becomes
\begin{gather}
| \hat{H} E \rangle = U | E \rangle .
\end{gather}
This single solution that arises when $N$ is even and $U\ne0$ must be added to
the eigenstates obtained using the Bethe anzats to form a complete set of
solutions.
\end{enumerate}
Taking this considerations into account, the total number of independent
solutions turns out to be $N(N-1)/2$, regardless of the parity of $N$. This is
the correct number of solutions in a system where two fermions can access $N$
single-body states.
%
%
%
%
%
%
%
%
%
%
%
%
%
%
%
%
%
%
\end{document}